\documentclass[aps,prb,twocolumn,superscriptaddress,floatfix,amsmath,longbibliography,10pt]{revtex4-2} 

\usepackage[normalem]{ulem}
\usepackage{bm} 
\usepackage{amssymb}
\usepackage{amsfonts} 
\usepackage{amsmath} 

\usepackage{graphicx} 

\usepackage{xcolor} 
 
\usepackage{siunitx}
\DeclareSIUnit{\dBm}{dBm}
\DeclareSIUnit{\Oe}{Oe}

\usepackage[pdftex,colorlinks=true,allcolors=blue,breaklinks=true]{hyperref}  

\usepackage[utf8]{inputenc} 
\usepackage{textgreek}
 
\definecolor{g-blue}{rgb}{0.83,0.95,1}
\definecolor{g-yellow}{rgb}{1,1,0.7}
\definecolor{g-green}{rgb}{0.9,1,0.9}
\definecolor{green}{rgb}{0,0.6,0}
\definecolor{cyan}{rgb}{0,0.7,0.7}
\definecolor{black}{rgb}{0,0,0}
\definecolor{grey}{rgb}{0.4,0.4,0.4}
\definecolor{nature-blue}{rgb}{0.0,0.200,0.500}

\def \ed {\end{document}}
\def\Fbox#1{\vskip1ex\hbox to 8.5cm{\hfil\fboxsep0.3cm\fbox{%
		\parbox{8.0cm}{#1}}\hfil}\vskip1ex\noindent}  

\let \= \equiv  
\let\*\cdot 
\let\^\widehat 
\let\-\overline

\def\1{\bm1}

\def\<{\left\langle}    \def\>{\right\rangle}
\def\[ {\left[}         \def\]{\right]}

\begin{document}


\title{Enhancement of magnon flux toward a Bose--Einstein condensate}


\author{Franziska Kühn}
\affiliation{Department of Physics and Research Center OPTIMAS, RPTU University Kaiserslautern-Landau, 67663 Kaiserslautern, Germany}

\author{Matthias R. Schweizer}
 \affiliation{Department of Physics and Research Center OPTIMAS, RPTU University Kaiserslautern-Landau, 67663 Kaiserslautern, Germany}

\author{Tamara Azevedo}
 \affiliation{Department of Physics and Research Center OPTIMAS, RPTU University Kaiserslautern-Landau, 67663 Kaiserslautern, Germany}

\author{Vitaliy I. Vasyuchka}
 \affiliation{Department of Physics and Research Center OPTIMAS, RPTU University Kaiserslautern-Landau, 67663 Kaiserslautern, Germany}

\author{Georg von Freymann}
 \affiliation{Department of Physics and Research Center OPTIMAS, RPTU University Kaiserslautern-Landau, 67663 Kaiserslautern, Germany}
 \affiliation{Fraunhofer Institute for Industrial Mathematics ITWM, 67663 Kaiserslautern, Germany}

\author{Victor S. L'vov}
 \affiliation{The Weizmann Institute of Science, Rehovot, 76100, Israel}
 \email{Victor.Lvov@gmail.com}

 \author{Burkard Hillebrands}
 \affiliation{Department of Physics and Research Center OPTIMAS, RPTU University Kaiserslautern-Landau, 67663 Kaiserslautern, Germany}

\author{Alexander A. Serga}
 \email{serha@rptu.de}
 \affiliation{Department of Physics and Research Center OPTIMAS, RPTU University Kaiserslautern-Landau, 67663 Kaiserslautern, Germany}

\date{\today}
\begin{abstract}  
We present a combined theoretical and experimental study of angle-dependent parametric pumping of magnons in Yttrium Iron Garnet films, with a focus on the mechanisms that transfer parametrically injected magnons toward the spectral minimum where Bose--Einstein condensation occurs. Using a classical Hamiltonian formalism, we analyze the threshold conditions for parametric instability as a function of the angle $\alpha_\mathrm{p}$ between the microwave pumping field $\bm h_\mathrm{p}(t)$ and the external magnetic field $\bm H_\mathrm{ext}$, continuously tracing the transition between parallel (for $\alpha_\mathrm{p}=0$) and transverse pumping (for $\alpha_\mathrm{p}=\pi/2$). We also describe two competing four-magnon scattering mechanisms that transfer parametric magnons toward the bottom of their frequency spectrum: The step-by-step Kolmogorov--Zakharov cascade, which is allowed for all values of $H_\mathrm{ext}$, and the kinetic instability mechanisms that provide a much more efficient single-step channel in transferring magnons directly to the lowest-energy states, but occurs within specific regions of $\alpha_\mathrm{p}$ and $H_\mathrm{ext}$ where the conservation laws permit it. 
In the experimental part, we employ microfocused Brillouin light scattering spectroscopy in combination with a vector magnet, allowing for angle-resolved mapping of the magnon population spectrum under controlled pumping angle $\alpha_\mathrm{p}$. We observe that transverse pumping, although characterized by a higher instability threshold, yields a markedly stronger population at the spectral minimum compared to parallel pumping. These observations demonstrate that the kinetic instability channel plays a dominant role in transferring magnons to the spectral minimum under such conditions. These results reveal the crucial role of pumping geometry in shaping the magnon distribution and provide guidelines for optimizing the flux of magnons into the condensate, thereby advancing the control of magnon Bose–Einstein condensation in magnetic insulators.
 \end{abstract}
\maketitle

\section{\label{s:intro}Introduction}
Since the discovery of the room-temperature Bose--Einstein condensate (BEC) \cite{Kalafati1991, Melkov1994, Demokritov2006} in a magnon gas overpopulated by electromagnetic parametric pumping applied to a single-crystal ferrimagnetic film of Yttrium Iron Garnet (Y$_3$Fe$_5$O$_{12}$, YIG) \cite{Cherepanov1993}, intensive research has elucidated the main mechanisms behind the formation of this spin state \cite{Demidov2008, Serga2014, Bozhko2015, Kreil2018, Lvov2024}, confirmed its temporal, spatial, and phase coherence \cite{Rezende2009, NowikBoltyk2012, Dzyapko2016, Noack2021, Koster_arxiv.2507.16862}, and revealed its relation to the simultaneous accumulation of quasiparticles in regions of magnon-phonon hybridization \cite{Bozhko2017, Frey2021, Hahn2022}. Furthermore, phenomena known from other Bose condensates, such as supercurrents \cite{Bozhko2016, Mihalceanu2019, Schweizer2022, Schweizer2024}, Josephson oscillations \cite{Kreil2018}, second sound \cite{Tiberkevich2019}, and Bogoliubov waves \cite{Bozhko2019}, have also been observed.

These advances have paved the way for further in-depth studies of fundamental topics, including the nonlinear dynamics of the magnon condensate, its stability and collapse \cite{Dzyapko2017, Borisenko2020, Frostad2024}, and the formation of vortex and domain structures \cite{NowikBoltyk2012, Bugrij2013, Sugakov2016, Mohseni2022}. 
They have also raised the prospect of practical applications, such as information transmission and processing in magnon circuits \cite{Mohseni2022}, energy harvesting via the conversion of chaotic spin-system excitations into coherent microwave signals \cite{Dzyapko2008}, emission of short-wavelength propagating spin waves (magnons) \cite{NowikBoltyk2019, Borisenko2020}, amplification of externally excited spin waves by the spectral flux of thermalizing magnons \cite{Breitbach2023}, and stabilization of spatially localized spin-wave bullets coexisting with the BEC \cite{Schneider2021_bullet}.

In both fundamental research and engineering applications, the population density of the magnon BEC achievable in experiments plays an essential role. 
Because the number of magnons, the quanta of collective spin oscillations, is temperature dependent, a magnon BEC (as well as BECs of other similar quasiparticles) cannot be achieved via slow, quasi-equilibrium cooling of a sample. Instead, external injection beyond thermodynamic equilibrium is required, with the BEC density determined by the number of injected magnons and their thermalization.
Although various injection mechanisms have been demonstrated---including rapid cooling of the magnetic sample \cite{Schneider2020} and spintronic magnon injection via the spin-Seebeck effect \cite{Safranski2017} or the combined action of the spin-Hall and spin-transfer torque effects \cite{Schneider2021, Divinskiy2021}---the most reliable, efficient, and widely used approach remains parametric pumping of magnons by an external electromagnetic field (see, e.g., \cite{Lvov2023, Lvov2024, Melkov2025}), which to date enables the highest magnon condensate densities to be achieved. 

The aim of this work is to systematically investigate the influence of the pumping geometry on the efficiency of magnon overpopulation, with the ultimate goal of identifying configurations that maximize the magnon flux into the low-energy region of the spectrum. 

The paper is organized as follows. 
In Section~\ref{s:exit}, we analytically describe the methods of parametric injection of magnons in magnetically ordered dielectrics (Sec.\,\ref{ss:methods} and \ref{ss:Ham-p}) and the mechanisms responsible for guiding these magnons toward the spectral minimum (Sec.\,\ref{ss:scattering}). This redistribution is a key prerequisite for the formation of magnon Bose--Einstein condensates.

Section~\ref{sec:exp} describes the experimental study, which relies on microfocused Brillouin light scattering (BLS) spectroscopy \cite{Sebastian2015} in combination with a parametric pumping circuit and a vector magnet that enables continuous in-plane rotation of the external magnetic field with respect to the orientation of the microwave pumping field \cite{Schweizer2023}. This approach allows direct access to the parametrically populated magnon spectrum under controlled pumping geometries, covering a broad range of frequencies and wavevectors with considerable sensitivity to low-energy states near the spectral minimum. Section~\ref{ss:setup} details the experimental arrangement, Sec.~\ref{ss:thresholds} presents the threshold behavior of parametric instability for different pumping angles, and Sec.~\ref{ss:bottom} reports angle-resolved BLS measurements that reveal the redistribution of magnons toward the spectral minimum.

Finally, Sec.~\ref{s:sum} summarizes and discusses the main findings of our research.  
We demonstrate a pronounced dependence of the threshold for parametric pumping on the angle $\alpha_\mathrm{p}$ between the alternating magnetic field $\bm{h}(t) \propto \exp(-i \omega_{\rm p} t)$ and the static external magnetic field $\bm{H}_\mathrm{ext}$.  
Our results reveal that the kinetic-instability mechanism, which transfers parametrically excited magnons to the bottom of their spectrum in a single step, is generally more efficient than the Kolmogorov--Zakharov step-by-step cascade process described below.   

Unexpectedly, the total population of low-frequency magnons is higher at those angles $\alpha_\mathrm{p}$ where parametric excitation itself is less efficient.  
This behavior originates from the enhanced effectiveness of the kinetic-instability channel under these conditions.  
The ability to control the dominant scattering processes by varying $\alpha_\mathrm{p}$, and to enhance the magnon flux toward the Bose--Einstein condensate under perpendicular-pumping conditions, enables the formation of steady-state dense magnon condensates and paves the way for systematic studies of their nonlinear dynamics. 

\section{\label{s:exit}Excitation and scattering of parametric magnons}
This section presents the theoretical framework for the parametric pumping of magnons and the mechanisms governing their subsequent thermalization toward the spectral minimum.
Section~\ref{ss:methods} introduces the methods of parametric excitation of magnons. Section~\ref{ss:Ham-p} formulates the Hamiltonians of parametric pumping for parallel (\ref{sss:par}), transverse (\ref{sss:trans}), and oblique (\ref{sss:obl}) geometries; the underlying Hamiltonian formalism is outlined in Appendix~\ref{A}. The analysis of the corresponding Hamiltonian equations of motion in Sec.~\ref{ss:mot} yields the thresholds of parametric instability for each geometry. Section~\ref{ss:scattering} examines, within the framework of a four-magnon kinetic equation, the dominant scattering channels that transfer magnons toward the spectral minimum $\omega_{\rm min}$. These include the Kolmogorov--Zakharov step-by-step cascade (Sec.~\ref{sss:Cascade}) and the kinetic instability phenomenon (Sec.~\ref{sss:KI}), which transfers magnons directly from $\omega_{\bm k}\!\approx\!\omega_{\rm p}/2$ to $\omega_{\rm min}$ in a single step. Finally, Section~\ref {ss:concl} summarizes the theoretical results and highlights open questions concerning the relationship between parametric thresholds and scattering-driven thermalization, which call for experimental investigation.

\subsection{\label{ss:methods}Methods of parametric excitations}
The most effective method for injecting magnons into selected regions of the spin-wave spectrum in magnetodielectric crystals is parametric excitation by an external electromagnetic field.
This specifically concerns selective injection, since simply raising the temperature of the sample can, of course, increase the number of thermal magnons to extraordinary levels \cite{Schneider2020}, but in such a case, any targeted population of particular spectral regions cannot be achieved.

Another widely used approach is ferromagnetic resonance, i.e., the excitation of magnetic moment precession by the direct action of an external torque.
However, as is well known from general oscillation theory, the maximum amplitude of excitation in this case is limited by the damping of the oscillatory system \cite{Landau1976Mechanics}.

In contrast, under parametric pumping, a different scenario emerges: parametric instability---seen as an exponential growth of oscillation amplitude---arises once damping is compensated by an external energy influx, which, like the energy losses, is proportional to the mean square of the amplitude, i.e., magnon number density. 
Under these conditions, the maximum achievable density of parametrically injected magnons is ultimately constrained by various nonlinear mechanisms \cite{Lvov1993}.

In addition to these amplitude-related constraints, ferromagnetic resonance is also subject to spatial limitations. Specifically, the shortest spin-wave wavelength that can be excited in this process is on the order of the size of the region where the alternating magnetic field is localized. Parametric pumping, by contrast, is free from this restriction: it enables the excitation of magnons with submicron wavelengths even when the driving field is spatially homogeneous.

The first method of parametric injection was proposed in 1957 by Suhl \cite{Suhl1957}. This scheme, known as \textit{transverse pumping of spin waves} and later referred to as Suhl’s first-order instability, relies on the decay of the homogeneous magnetization precession with frequency $\omega_{\rm p}$ into a pair of spin waves (magnons) with opposite wavevectors $\bm k$ and $-\bm k$. In the quasiparticle picture, a magnon of frequency $\omega_{\rm p}$ and approximately zero wavevector splits into two magnons with wavevectors $\bm k$ and $-\bm k$. Both magnons carry the same frequency, $\omega_{\bm k} = \omega_{-\bm k} = \omega_{\rm p}/2$.
If this process is prohibited by energy and momentum conservation, Suhl's second-order instability can occur: two magnons associated with a resonantly or off-resonantly driven homogeneous precession at $\omega_{\rm p}$ decay into a pair of magnons with frequencies $\omega_{\rm p}=\omega_{\pm \bm k}$ and wavevectors $\pm \bm k$.

In 1962, Schl\"{o}mann proposed an alternative scheme, referred to as \textit{parallel pumping} \cite{Schloemann1962}. In this case, the decay instability occurs not in the precession but in an externally applied, nearly homogeneous magnetic field with frequency $\omega_{\rm p}$, whose photons again split into pairs of magnons with wavevectors $\pm \bm k$ and frequency 
$\omega_{\bm k} = \omega_{-\bm k} = \omega_{\rm p}/2$.

In Schl\"{o}mann’s parallel pumping process, the alternating magnetic field $\bm h(t) = \bm h^\|(t)$ is aligned with the equilibrium magnetization $\bm M_0$, while in Suhl’s transverse pumping process $\bm h(t) = \bm h ^\perp(t)$ is oriented perpendicular to $\bm M_0$. In this work, we also consider intermediate configurations, referred to as \textit{oblique pumping}, where the angle $\alpha_\mathrm{p}$ between $\bm h(t)$ and 
$\bm M_0$ is arbitrary. The corresponding threshold conditions for oblique pumping will be analyzed in the next section within the framework of the classical Hamiltonian formalism, briefly outlined in Appendix~\ref{A}.

\subsection{\label{ss:Ham-p}Hamiltonians of parametric pumping}
Within the framework of the classical Hamiltonian formalism, the equation of motion for small spin-wave amplitudes $c(\bm k, t)\equiv c_{\bm k}$ can be written in canonical form 
\begin{equation}\label{Heq1}
   i\Big ( \frac{\partial }{\partial t} + \gamma_{\bm k} \Big ) c_{\bm k} =\frac{\delta \cal H}{\delta c^*_{\bm k}}\,, 
\end{equation}
where $\gamma_{\bm k}$ is the wavevector-dependent phenomenological damping frequency of the spin waves \cite{Zakharov1992,Zakharov2025,Nazarenko2011}. Hereafter, $\mathcal{H}$ represents the Hamiltonian of the system, which corresponds to its energy expressed via the amplitudes $c_{\bm k}$ and $c_{\bm k}^*$ for all wavevectors. Its expansion for small amplitudes starts with the quadratic term: $ \mathcal{H}= \mathcal{H}_2 + \dots$, where 
\begin{eqnarray}\label{1} 
   \mathcal{H}_2 &=& \int \omega_{\bm k}c_{\bm k} c_{\bm k}^* d^3 k 
\end{eqnarray}
describes non-interacting spin waves. 

\subsubsection{\label{sss:par}Parallel pumping} 
As shown by Schl\"{o}mann in Ref.\,\cite{Schloemann1962}, the longitudinal part of the Zeeman energy $-h^\|(t)\, M_z(\mathbf{r},t)$ adds an additional term ${\cal H}^\|_{\rm p}$ to the total Hamiltonian ${\cal H}$. Here, $h^\|(t)=2\operatorname{Re} \{ h^\| \exp (-i \omega_{\rm p}t) \}$ and $M_z$ is the $z$-component of the magnetization $\bm M$ precessing about the static magnetic field $\bm H$ directed along the $z$-axis. 
For more details, see, e.g., Eqs.\,(4.3.17-4.3.20) in Ref.\,\cite{Lvov1993}. 

The pumping Hamiltonian ${\cal H}^\|_{\rm p}$ is expressed in the following form:
\begin{subequations}\label{Hp-par}
  \begin{align}\label{Hp-parA}
    {\cal H}^\|_{\rm p}= &\frac 12 \int \big [ h ^\| \exp (-i \omega_{\rm p}t ) V_{\bm k}^\| c_{\bm k}^*c_{-\bm k}^* + \mbox{c.c.} \big]\,,\\ 
       V_{\bm k}^\| = & {\cal V}_{\bm k}^\|\sin^2 \theta_{\bm k} 
       \,,   \label{Hp-parB}
    \quad   {\cal V}_{\bm k}^\| =   \frac {g \omega_{_{M}}}{2 \omega_{\bm k}}= \frac{g   \omega_{_{M}}}{\omega_{\rm p}} \ .
  \end{align}
\end{subequations} 
Hereafter, $\theta_{\bm k}$ represents the angle between $ \bm k $ and $ \bm M $, c.c. denotes complex conjugation, $\omega_{_{M}}=4\pi g M$, and $g$ is the gyromagnetic ratio. 

\begin{figure}[!htbp]
\centerline{\includegraphics[width=1\columnwidth]{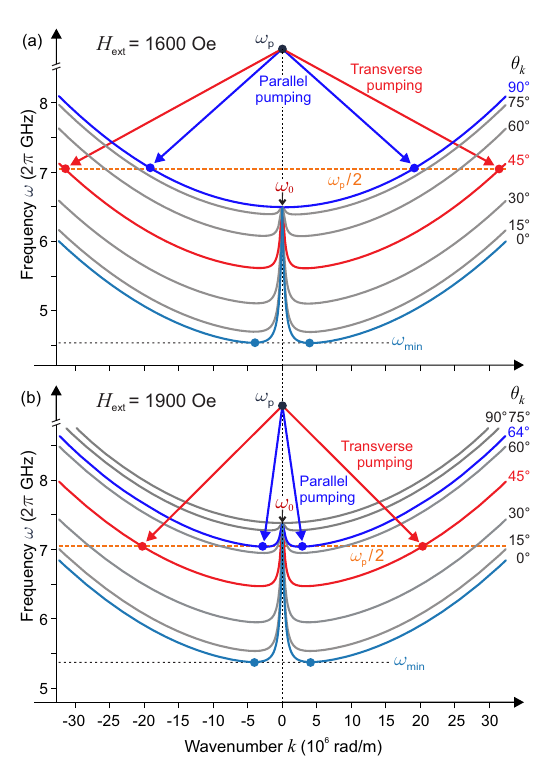}}
\caption{ 
Calculated dispersion relation for a \qty{6.7}{\micro\meter}-thick in-plane--magnetized YIG film at room temperature in external fields (a) $H_\mathrm{ext} = \qty{1600}{\Oe} < H_\mathrm{crit}$ and (b) $H_\mathrm{ext} = \qty{1900}{\Oe} > H_\mathrm{crit}$, where $H_\mathrm{crit}\!\approx\!\qty{1750}{\Oe}$. At $H_\mathrm{ext}=H_\mathrm{crit}$, the frequency of uniform precession $\omega_0$ (the minimum of the transverse branch with $\theta_{\bm k}=\pi/2$ at $k=0$) coincides with the frequency $\omega_\mathrm{p}/2$. Blue dots denote magnons excited by parallel pumping, where the minimal threshold is reached at $\theta_{\bm k}=\pi/2$ for $H_\mathrm{ext} < H_\mathrm{crit}$ in (a) and shifts to $\theta_{\bm k}<\pi/2$ for $H_\mathrm{ext} >H_\mathrm{crit}$ in (b). Red dots denote magnons excited by transverse pumping, with the minimal threshold at $\theta_{\bm k}\!\approx\!\pi/4$. The corresponding parametric processes are illustrated by blue and red arrows, representing the decay of a microwave photon and a magnon into two magnons, respectively.  
The cyan dots indicate two global minima of the magnon spectrum, which are located on the longitudinal dispersion branches with $\theta_{\bm k}=0$.
} 
\label{fig:Disp}
\end{figure}

The process of parallel pumping is illustrated in Fig.\,\ref{fig:Disp} by blue arrows. 
In Fig.\,\ref{fig:Disp}\,(a), in agreement with Eq.\,\eqref{7}, the excitation angle $\theta_{\bm k}$ of parametric waves with the minimal threshold equals $\pi/2$ when the external magnetic field $H_\mathrm{ext}$ is below the critical value $H_\mathrm{crit}$. 
However, when $H_\mathrm{ext}$ exceeds $H_\mathrm{crit}$, as shown in Fig.\,\ref{fig:Disp}\,(b), these processes become forbidden by the conservation laws, and the excitation angle tilts toward $\theta_{\bm k} < \pi/2$.

The physical origin of the parallel pumping Hamiltonian ${\cal H}^\|_{\rm p}$ describing the parametric excitation of the pair of spin waves by the longitudinal field $\bm h^\|$ can easily be illustrated by the following simple geometric consideration.
Due to the magnetic dipole interaction, the magnetization at any point precesses along an elliptic cone (formally it is revealed by the circular variables $b_{\bm k}$ so that, in order to diagonalize the Hamiltonian ${\cal H}_2$, Eq.\,\eqref{H2}, the Bogoliubov $u-v$-transformation \eqref{uvA} is necessary).
Since the length of the vector $\bm M$ remains constant, the base of the cone is not flat, which results in the appearance of an oscillating longitudinal component ($z$-component) of the vector $\bm M$, varying at twice the precession frequency, $2 \omega_{\bm k}$. Clearly, those waves can be excited by a magnetic field of frequency $2 \omega_{\bm k}$ applied along $ \hat {\bm z}$ direction. \looseness=-1

\subsubsection{\label{sss:trans}Transverse pumping} 
Usually, the homogeneous precession is excited by a magnetic field $\bm h^\perp$ directed transverse to the constant magnetic field $\bm H$ (and accordingly to the stationary magnetization $\bm M_0$). The Hamiltonian of this interaction is due to the transverse part of the Zeeman energy $-\bm h^\perp \cdot \bm M^\perp$ and has the form 
\begin{equation}\label{4}
{\cal H}_{\rm p}= - \sqrt \frac{g M_0}2[h^\perp c_0 + \mbox{c.c}]\,, \quad h^\perp=h_x+i h_y\ . \end{equation} 
This Hamiltonian describes the well-known phenomenon of the homogeneous ferromagnetic resonance
with amplitude $c_0$. For the right-hand polarization \cite{Lvov1993} 
\begin{align}\label{5}
    c_0(t)=   c_0 \exp (-i \omega _{\rm p }t)\,, \quad
    c_0= \frac{\sqrt{g\, M_0/2}\, h^\perp }{\omega _{\rm p}-\omega_0 + i \gamma _0} \ .
\end{align} 
Here, $\omega_0$ represents the eigenfrequency of homogeneous precession, while $\gamma_0$ denotes its damping frequency. Equation \eqref{5} is also valid for the linear polarization near the resonance frequency.  
For a more detailed derivation, including the case of linear polarization, see, e.g., Eqs.\,(4.3.7--4.3.10) in Ref.\,\cite{Lvov1993}.

The homogeneous precession of magnetization decays into a pair of magnons with wavevectors $\bm{k}$ and $-\bm{k}$, both having a frequency of $\omega_{\bm{k}} = \omega_{-\bm{k}} = \omega_{\text{p}}/2$. This process is described by specific terms in the three-wave interaction Hamiltonian 
\begin{equation}\label{6}
    {\cal H}_3^ \perp= \frac12 \int \big [ V_{0, \bm k, -\bm k}\ c_0(t) c_{\bm k}^* c _{-\bm k}^*+ \mbox{c.c} \big ] d^3 k\,, 
\end{equation}
with the interaction amplitudes\,\cite{Lvov1993}
\begin{align}
    \begin{split}\label{7} 
        V_{0, \bm k, -\bm k} =& {\cal V}_{ \bm k }^\perp \sin 2\theta_{\bm k}\,, \\ 
        {\cal V}_{ \bm k }^\perp  =& \frac{\pi g }{\omega_{\rm p}}\sqrt{2g M_0} \\
        \times &\Big [ \omega_{\rm p} + \omega_{_{M}}\sin^2\theta_{\bm k}+ \sqrt{\omega_{\rm p}^2+ \omega_{_{M}}^2\sin\theta_{\bm k}}\,\Big]_1\ .
    \end{split}
\end{align}
Substituting $c_0(t)$ from \eqref{5} into Eq.\,\eqref{6} for ${\cal H}_3^ \perp$,
we obtain the following expression for the effective Hamiltonian of the transverse pumping 
\begin{subequations}\label{8}
    \begin{align}\label{8A}
        {\cal H}^\perp_{\rm p}= &\frac 12 \int \big [ h ^\perp  \exp (-i \omega_{\rm p}t ) V_{\bm k}^\perp c_{\bm k}^*c_{-\bm k}^* + \mbox{c.c.} \big] d\bm k\,,  \\ \label{8B}
        V_{\bm k}^\perp =&  g \sin 2\theta_{\bm k}  \frac{\omega_{_{M}}}{  \omega_{\rm p} \, } \frac{[\omega_{\rm p}+ \dots]_1}{4(\omega_{\rm p}-\omega_0 + i \gamma_0)}\,,
    \end{align}  
\end{subequations}
where $ [\omega_{\rm p}+ \dots]_1$ is the expression in parentheses in \eqref{7}. 
A schematic representation of the transverse pumping process is shown in Fig.\,\ref{fig:Disp} by red arrows. 
It can be seen that the excitation angle $\theta_{\bm k}$ depends only weakly on the relation between $H_\mathrm{ext}$ and $H_\mathrm{crit}$.

\subsubsection{\label{sss:obl}Oblique pumping} 
In the case of oblique pumping, the Hamiltonian comprises the sum of the parallel and transverse contributions: 
 \begin{subequations}\label{9}
    \begin{align} \label{9A}
        {\cal H}^\angle_{\rm p}=  \frac 12 \int (\bm h \cdot \bm V_{\bm k} )& \exp (-i \omega_{\rm p}t )  c_{\bm k}^*c_{-\bm k}^* + \mbox{c.c.} \big]d^3 k\,,  \\ \label{9B}
        (\bm h \cdot \bm V_{\bm k})&= \, h ^\| V_{\bm k}^\| + h ^\perp V_{\bm k}^\perp \ . 
    \end{align} 
\end{subequations} 

\subsection{\label{ss:mot}Threshold of parametric instability} 
Motion equation \eqref{Heq1} with the Hamiltonian ${\cal H}= {\cal H}_2+{\cal H}_{\rm p}^\angle$, taken  from \eqref{1} and \eqref{9}, has the form:
\begin{align}
    \begin{split}\label{10}
        \Big [ \frac{\partial }{\partial t} +& \gamma_{\bm k}+ i \omega_{\bm k}\Big ]c_{\bm k}(t)\\ 
        +& i (\bm h \cdot \bm V_{\bm k})\exp (i \omega_{\rm p}t)c_{-\bm k}^*(t)=  0\,, \\
        \Big [ \frac{\partial }{\partial t} +&  \gamma_{\bm k}- i \omega_{\bm k}\Big ]c_{-\bm k}^*(t)\\ 
        -& i (\bm h \cdot \bm V_{\bm k})^*\exp (-i \omega_{\rm p}t)c_{ \bm k} (t)=  0\ .
    \end{split} 
\end{align}
The system of linear homogeneous Eqs.\,\eqref{10} has non-trivial solutions 
\begin{align}
    \begin{split}\label{11}  
        \hskip - .2cm c_{\bm k}(t)\propto & \exp [(\nu -i  \frac{\omega_{\rm p}}2 )t]\,,  \ c_{-\bm k}^*(t)\propto \exp [(\nu +i \frac{\omega_{\rm p}}2 )t]\,,
        \\ 
        \mbox{with} &\quad  \nu_{\bm k} = 
        -\gamma_{\bm k} \pm  \sqrt { |{\bm h \cdot \bm V_{\bm k} |^2} - \Big( \frac{\omega_{\rm p}}2 - \omega_{\bm k}\Big )^2  }\ . 
    \end{split} 
\end{align}

The threshold for parametric instability ($\nu_{\bm k}=0$), which corresponds to the excitation of spin waves with $\omega_{\pm \bm k} = \omega_{\rm p}/2$, is defined by
\begin{equation} \label{12}
  | \bm h \cdot \bm V_{\bm k} | = \gamma_{\bm k}\ .
\end{equation}
As a result, as $h$ increases, the pairs of magnons for which the ratio $| \bm h \cdot \bm V_{\bm k} |/\gamma_{\bm k}$ is minimized are excited first. 

For the parallel pumping, this occurs for $\theta_{\bm k}= \pi /2$ in the simple case when $ \gamma_{\bm k}$ is independent of the direction of $\bm k$. Under these conditions, $ \bm{h} \cdot \bm{V}_{\bm{k}} = h^\| V_{\bm{k}}^\| \propto \theta \sin^2 \theta_{\bm{k}} $, as shown in Eq.\,\eqref{Hp-parB}. 
Their threshold is given by
\begin{equation}\label{13} 
    g h^\|_{\rm th}= \frac{\omega_{\rm p}}{\omega_{_{M}}}\gamma^\|\,,
\end{equation}
where $\gamma^\|$ represents the damping frequency of magnons at the minimum threshold for parallel pumping, 
see Eqs.\,\eqref{12} and \eqref{Hp-parB}. 
In our experiment, where $\omega_{\rm p}/(2\pi) = \qty{14.094}{\giga\hertz}$ and $\omega_{M}/(2\pi) \approx \qty{4.9}{\giga\hertz}$ (see Sec.~\ref{sec:exp}), this corresponds to about $2.9 \gamma^\|$.

For the transverse pumping case $ \bm h \cdot \bm V_{\bm k} =h^\perp V_{\bm k}^\perp $. According to Eqs.\,\eqref{7} and \eqref{8B} the most important part of the angular dependence is $\sin2 \theta_{\bm k}$. Therefore, for a simple case of angular independent $\gamma_{k}$, the minimal threshold corresponds to the excitation on the resonance surface of two meridians $ \theta_{\bm k}=\theta_{\rm min}\approx \pi / 4$ and $\theta_{\bm k} = \pi/2-\theta_{\rm min}$ with the threshold
\begin{align}\label{14}
    g h^\perp_{\rm th} \approx\frac{\omega_{\rm p}}{\omega_{_{M}}}\gamma _k
    \frac{ 4 \sqrt{  ( \omega_{\rm p}-\omega_0 )^2 + \gamma_0^2 } }{[\omega_{\rm p}+\dots]_1} \ . 
 \end{align}
Under our experimental conditions, with $\omega_{0}/(2\pi) \approx \qty{6.929}{\giga\hertz}$ and $\omega_{\rm p}/(2\pi) \approx \qty{14.094}{\giga\hertz}$, we obtain $\theta_{\rm min}\approx \pi/4 + 0.05$ and $g h^\perp_{\rm th}\approx 2.4\,\gamma^\perp$. Bearing in mind that the wavenumber of parametric magnons excited by transverse pumping is usually larger than that by parallel pumping, we expect $\gamma^\perp \gtrsim \gamma^\|$. We see that far from the ferromagnetic resonance (which occurs when $\omega_{\rm p} = \omega_0$), the threshold for transverse pumping is close in magnitude to that of parallel pumping.

Another distinction between the two pumping regimes lies in the region of the excited parametric waves: for parallel pumping, the angle $\theta^{\|}_{\bm k}$ is approximately $\pi/2$, whereas for transverse pumping, $\theta^{\perp}_{\bm k}$ is about $\pi/4$. 

In the general case of oblique pumping, the pumping amplitude $\bm h \cdot \bm V_{\bm k}$, as shown in Eq.\,\eqref{9B}, consists of two components: one corresponding to parallel pumping and the other to transverse pumping. The relative contributions of these components depend on the angle $\alpha_\mathrm{p}$ between the external constant magnetic field $\bm H_\mathrm{ext}$, and the alternating field $\bm h$
\begin{equation}\label{oblique}
    \bm h \cdot \bm V_{\bm k} = h \big[ V_{\bm k}^\|\cos\alpha_\mathrm{p}+ V_{\bm k}^\perp \sin\alpha_\mathrm{p} \big ]\ .
\end{equation}
The threshold for oblique pumping is determined by the condition $ | \bm{h} \cdot \bm{V}_{\bm{k}} | \sim \gamma_{\bm{k}}^\angle $, where $ \gamma_{\bm{k}}^\angle $ represents the magnon damping associated with the excitation angle $ \theta_{\bm{k}}^\angle $. This angle ranges between $ \dfrac{\pi}{4} $ and $ \dfrac{\pi}{2} $. Furthermore, under our conditions, the threshold for oblique pumping will vary with $\alpha_\mathrm{p}$ between $2.4 \gamma_{\bm{k}}$ and  $2.9 \gamma_{\bm{k}}$ (even though $\gamma_{\bm{k}}$ will be independent of $\alpha_\mathrm{p}$).

The present analysis was performed for spatially homogeneous pumping in an isotropic, unbounded magnetic medium. In contrast, most contemporary experiments, including our current work, employ thin-film samples with spatially localized pumping, for example, using microstrip resonators. In such geometries, parametrically injected magnons can escape from the pumping region. 
A change in the direction of the parametric magnons’ wavevector, arising from a transition between parallel and perpendicular pumping---either through rotation of the external magnetic field, as in our work, or by varying its magnitude, as in Ref.\,\cite{Neumann2009_2}---can enhance their losses, thus raising the threshold of parametric instability.

\subsection{\label{ss:scattering}Scattering of parametrically injected magnons toward the BEC}
 
The theory of weak wave turbulence \cite{Zakharov1992, Zakharov2025, Nazarenko2011} provides a consistent framework for describing weakly interacting quasiparticles in terms of the quasiparticle occupation number $n(\bm{k},t) \equiv n_{\bm{k}}$, defined as follows:
\begin{equation}\label{15}
    (2\pi)^3 \delta({\bm k-\bm k'})n_{\bm k}= \langle c_{\bm k} c_{\bm k'}^*\rangle \ .
\end{equation}
Here $\langle \dots\rangle $ denotes a ``proper'' averaging. 
It may include spatial averaging in a uniformly distributed scenario, time averaging in a steady situation, ensemble averaging in theoretical analyses, numerical simulations, or multiple experimental measurements. Without going into a detailed mathematical analysis, we assume that turbulent statistics are ergodic and that all these types of averaging are equivalent. 

From a physical perspective, a system of weakly interacting magnons can be conceptualized as a gas of magnons (see, e.g., Ref.\,\cite{Kaganov1997}). Due to intensive parametric pumping, its distribution can significantly exceed the thermodynamic equilibrium level and evolve toward Bose--Einstein condensation. This evolution can be described by the kinetic equation (KE)
\begin{equation}\label{16A}
    \frac{\partial n_{\bm k}}{\partial t} = \mbox{St}_{\bm k} { - 2 \gamma_{\bm k}n_{\bm k} }\,, 
\end{equation}
where $ \gamma_{\bm k}$ is the phenomenologically introduced damping frequency in Eq.\,\eqref{Heq1}, originated from processes not included in $ \mbox{St}_{\bm k}$. 
The collision integral $\mathrm{St}(\bm k, t) \equiv \mathrm{St}_{\bm k}$ can be obtained in various ways \cite{Zakharov1992, Nazarenko2011}, for example, from the Golden Rule, which is widely used in quantum mechanics\,\cite{Landau1970}.
\looseness=-1

For magnons with small wavenumbers $k$, in systems with a gapped dispersion law---i.e., with a non-zero bottom frequency $\omega_{\rm min}$---three-magnon processes are forbidden when the frequencies of the scattered magnons fall into a spectral gap with no available eigenstates. For an isotropic parabolic dispersion described by $\omega_k=\omega_{\rm min} + A k^2$, this condition holds true if $\omega_k < 3\, \omega_{\rm min}$ \cite{Lvov1993}. 
\looseness=-1

\subsubsection{\label{sss:Cascade}Kolmogorov--Zakharov scattering cascade}
In the absence of three-magnon scattering, $\mathrm{St}_{\bm k}$ is dominated by four-magnon processes
\begin{subequations}\label{16}
\begin{equation}\label{16B}
    \omega_{\bm k}+\omega_{\bm 1}=\omega_{\bm 2}+\omega_{\bm 3}, \quad \bm k + \bm 1 = \bm 2 + \bm 3\,, 
\end{equation} 
with the interaction Hamiltonian
\begin{align}
    \begin{split}\label{16C}
        {\cal H}_4=& \frac14 \int d \bm k d \bm k_1 d \bm k_2 d \bm k_3  T_{\bm k \bm 1}^{\bm 2\bm 3} c_{\bm k}^* c_{\bm 1}^* c_{\bm 2} c_{\bm 3}\\ &\times\delta(\bm k + \bm 1 - \bm 2 - \bm 3)\ .
    \end{split}
\end{align} 
Here, $\omega_{\bm j}= \omega( \bm k_j)$ and $ T_{\bm k \bm 1}^{\bm 2\bm 3}$ is the interaction amplitude. The bold indices $\bm 1$, $\bm 2$, and $\bm 3$ denote the wavevectors $\bm k_1$, $\bm k_2$, and $\bm k_3$, respectively. In this case, the collision integral reads \cite{Zakharov1992,Zakharov2025,Nazarenko2011}:
\begin{align}
    \begin{split}\label{16D}
        ^4\mathrm{St}_{\bm k}=& \frac{\pi}{4}\int d \bm k_1 d \bm k_2 d \bm k_3 \delta(\bm k + \bm 1 - \bm 2 - \bm 3)\\
        &\times \delta (\omega_{\bm k}+\omega_{\bm 1}-\omega_{\bm 2}-\omega_{\bm 3}) |T_{\bm k \bm 1}^{\bm 2\bm 3}|^2\\
        &\times [n_{\bm 2} n_{\bm 3} ( n_{\bm k} + n_{\bm 1}) - n_{\bm k} n_{\bm 1} ( n_{\bm 2}+n_{\bm 3} ]\ .
    \end{split}
\end{align} 
\end{subequations} 
The kinetic equation \eqref{16A} admits a stationary thermodynamic equilibrium solution, which is given by the Rayleigh--Jeans distribution as a high-temperature limit of the Bose--Einstein distribution: 
\begin{equation}\label{RJ}
    n_{\bm k}^{^{\rm RJ}} = \frac{T}{\omega_{\bm k} - \mu} \ .
\end{equation}
In this context, $T$ represents the temperature in energetic units, while $\mu$ denotes the chemical potential. The chemical potential $\mu$ is equal to zero in systems lacking particle-number conservation, such as magnon systems.
 
If $n_{\bm k}$ slightly deviates from $n_{\bm k}^{^{\rm RJ}}$, then the KE\,\eqref{16A} governs the exponential evolution of $n_{\bm k}$ toward $n_{\bm k}^{^{\rm RJ}}$:  
\begin{equation}\label{toward}
    n_{\bm k} - n_{\bm k}^{^{\rm RJ}} \propto \exp \big [-2 (\gamma_{\bm k} + ^{4}\!\!\gamma_{\bm k}) t \big ] \ .
\end{equation}
Here, the damping frequency $^{4}\gamma_{\bm k}$, originating from four-magnon scattering, is just a proportionality coefficient in Eq.\,\eqref{16D} in front of $-2n_{\bm k}$: 
\begin{align}
    \begin{split}\label{19}
        ^4\gamma_{\bm k}=& \frac{\pi}{8}\int d \bm k_1 d \bm k_2 d \bm k_3 \delta(\bm k + \bm 1 - \bm 2 - \bm 3)\\
        &\times  \delta (\omega_{\bm k}+\omega_{\bm 1}-\omega_{\bm 2}-\omega_{\bm 3}) |T_{\bm k \bm 1}^{\bm 2\bm 3}|^2\\
        &\times [   n_{\bm 1} ( n_{\bm 2}+n_{\bm 3}) - n_{\bm 2} n_{\bm 3}   ]\ .
  \end{split}
\end{align} 
When a restricted region of $\bm k$-space becomes strongly overpopulated ($n_{\bm k} \gg n_{\bm k}^{^{\rm RJ}}$), e.g., near $\omega_{\bm k} \approx \omega_{\rm p}/2$ under intense parametric pumping, the KE\,\eqref{16A} describes the resulting fluxes of energy and particle number out of this region.
For larger values of $k$, where the exchange interaction dominates over the dipole-dipole interaction, $\omega_{\bm k} \propto k^2$, and $T_{\bm k \bm 1}^{\bm 2 \bm 3}$ is a second-order homogeneous function of $k$: \looseness=-1 
\begin{equation}
    T_{\lambda \bm k, \lambda \bm 1}^{\lambda \bm 2, \lambda \bm 3} 
    = \lambda^2 T_{\bm k \bm 1}^{\bm 2 \bm 3} 
\end{equation}
(see, e.g., Ref.~\cite{Lvov1993}). As shown in Ref.~\cite{Lvov2024}, in this case, the energy flux is directed towards large $k$ and $\omega_{\bm k}$ (``direct energy cascade''), while the magnon number flux is directed towards small $k$ and $\omega_{\bm k}$ (``inverse particle cascade''). 

This statement is based on the analysis of the energy and quasiparticle number balance in the stationary, scale-invariant, isotropic situation adopting the fundamental relationship between the quasi-particle energy $\varepsilon(\omega)$ and their number $n(\omega)$ at a given frequency $\omega$\,: $ \varepsilon(\omega)=\omega n(\omega)$. 
One can assume that energy and quasiparticles are pumped around an intermediate frequency $\omega_{\rm p }$ (with pumping rates $W_{\rm p}^{^{\rm QP}}$ for quasiparticles and  $W_{\rm p}^{^{\rm E}} = \omega_{\rm p} W_{\rm p}^{^{\rm QP}}$ for energy) and dissipate at the final scattering regions: at low frequencies near the spectral bottom $\omega_{\rm min} < \omega_{\rm p}$ (with dissipation rates of quasi-particles $W_<^{^{\rm QP}}$ and of energy $W_<^{^{\rm E}} = \omega_{\rm min} W_<^{^{\rm QP}}$) and at high frequencies $\omega_\mathrm{max} \gg \omega_{\rm p}$ (with dissipation rates $W_>^{^{\rm QP}}$ and $W_>^{^{\rm E}} = \omega_{\rm max} W_>^{^{\rm QP}}$).
The particle-number and energy balances are then given by
\begin{subequations}\label{bal}
    \begin{align}\label{balA}
    W_{\rm p}^{^{\rm QP}} & = W_<^{^{\rm QP}} +  W_>^{^{\rm QP}}\,, \\ \label{balB}
        \begin{split}
            W_{\rm p}^{^{\rm E}} = \omega_{\rm p} W_{\rm p}^{^{\rm QP}} & =  W_<^{^{\rm E}}+  W_>^{^{\rm E}} \\ & =  \omega_{\rm min} W_<^{^{\rm QP}} + \omega_{\rm max} W_>^{^{\rm QP}}\ .   
        \end{split}
    \end{align}
\end{subequations}
In the considered case where $ \omega_{\rm min} < \omega_{\rm p} \ll \omega_{\rm max} $, it follows from Eqs.\,\eqref{bal} that most of the energy is dissipated in the high-frequency region. By contrast, quasiparticles predominantly decay in the low-frequency region. Consequently, the energy flux is directed toward high frequencies, while quasiparticles mainly flow toward low frequencies. 
In the theory of weak-wave turbulence, these fluxes are known as Kolmogorov--Zakharov cascades \cite{Zakharov2025}. It can further be shown that the inverse particle cascade 
is local in the sense of having a step-by-step nature: at each step, particles are transferred from a region $k_j$ to a compatible lower-$k$ region $k_{j+1} < k_j$ \cite{Nazarenko2011}. 

\subsubsection{\label{sss:KI}One-step scattering and the kinetic instability phenomenon}
There exists another channel of magnon transport from the overpopulated parametric region with $\omega_{\bm k} \approx \omega_{\rm p}/2$ directly down to the lowest frequency in the system, $\omega_{\rm min}$ \cite{Lavrinenko1981}. This channel originates from a specific four-magnon scattering process \eqref{16B}: two parametrically injected magnons with $\omega_{\bm k} \approx \omega_{\bm k_1} \approx \omega_{\rm p}/2$ merge, producing a secondary ``bottom'' magnon with $\omega_{\bm 2} \approx \omega_{\rm min} < \omega_{\rm p}/2$ and a secondary high-frequency ``top'' magnon with $\omega_{\bm 3} = \omega_{\rm p} - \omega_{\rm min} > \omega_{\rm p}/2$.

The physical origin of this channel can be clarified using Eq.\,\eqref{19}, which defines $^{4}\gamma_{\bm k}$.
It can be shown, as a general theorem, that we have $\gamma_{\bm k} > 0$ in thermodynamic equilibrium.
However, this is not necessarily valid far from equilibrium: in that case, the last term in the square brackets of Eq.\,\eqref{19}, proportional to $n_{\bm 2}n_{\bm 3}$, may dominate, rendering the overall expression in Eq.\,\eqref{19} for $^{4}\gamma_{\bm k}$ in the four-magnon scattering process \eqref{16B} negative. Following Ref.~\cite{Lvov2024}, we denote this contribution as $\Gamma^{^{\rm KI}}$ and estimate as 
\begin{align}
    \begin{split}\label{20}
        \Gamma^{^{\rm KI}}(k) \approx - & \int dk_1 dk_2 dk_3 k_1 k_2 k_3 |T_{\bm k \bm 1}^{\bm 2\bm 3}|^2\\ 
         \times & \delta(\omega_k + \omega_1-\omega_2 -\omega_3 )n^{\rm par}(k_2)n^{\rm par}(k_3)\\ 
        \approx & - \frac {  (\Omega_{_T}^{\rm par})^2 }{\omega _{\rm p}}\,, \quad \Omega_{_T}^{\rm par}= \langle |T_{\bm k \bm 1}^{\bm 2\bm 3}|^2 \rangle {\cal N}_{\rm par} \,,
    \end{split}
\end{align}
see Eqs.\,(36c) and (37) in Ref.\,\cite{Lvov2024}. Here, $ \langle |T_{\bm k \bm 1}^{\bm 2\bm 3}|^2\rangle $ is the mean square of the interaction amplitude and ${\cal N}_{\rm par} $ is the total number of parametric magnons. 
When the total number of parametric magnons ${\cal N}_{\rm par}$ is sufficiently large, the contribution $\Gamma^{^{\rm KI}}(k)$ can outweigh the other terms in $\gamma_{\bm k}$ arising from three-magnon interactions, scattering on defects and impurities. As a result, the total damping  
\begin{equation}
    \gamma_{\bm k}^{\rm tot} = \Gamma^{^{\rm KI}}(k) + \gamma_{\bm k}
\end{equation}
can become negative in certain regions of $\bm k$-space, leading to exponential growth of the secondary magnons,  
\begin{equation}
    {\cal N}_{\rm bot},\, {\cal N}_{\rm top} \propto \exp(-\nu_{\bm k} t)\ .
\end{equation}
This effect, first reported in Ref.\,\cite{Lavrinenko1981}, is known as \emph{kinetic instability} (KI).

\begin{figure}[t]
 \centerline{\includegraphics[width=1\columnwidth]{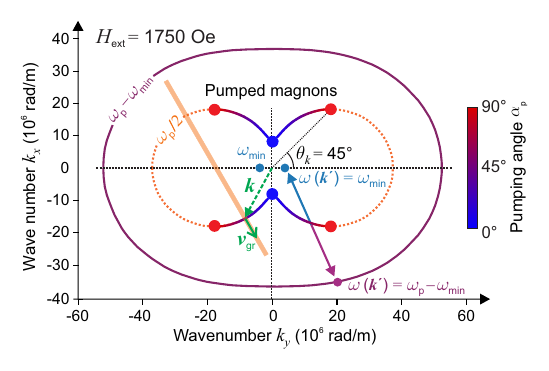}}
\caption{ 
Isofrequency contours of magnons in the $(k_x,k_y)$ plane for $H_\mathrm{ext} = \qty{1750}{\Oe}$ and $\omega_{\rm p}/(2\pi) = \qty{14.094}{\giga\hertz}$. Red dots indicate magnons parametrically excited by perpendicular pumping ($\alpha_\mathrm{p}=\qty{90}{\degree}$) with $\omega(\bm{k}_\mathrm{p}) = \omega_\mathrm{p}/2$, while blue dots correspond to magnons excited by parallel pumping ($\alpha_\mathrm{p}=\qty{0}{\degree}$). The red-to-blue segments of the magnon isofrequency contour $\omega_\mathrm{p}/2$ schematically indicate the spectral positions of parametrically excited magnons for different oblique-pumping angles (see the color scale for the pumping angle $\alpha_\mathrm{p}$). 
The dashed and solid green arrows represent the wavevector and the group velocity of parametrically excited magnons, respectively. In this schematic example ($\alpha_\mathrm{p}=\qty{30}{\degree}$), the latter is directed along the axis of the pumping resonator (depicted as the yellow stripe), while the wavevector is shown qualitatively to indicate the alignment of the group velocity with the resonator axis.
The double-headed cyan--magenta arrow illustrates the scattering of two pumped magnons into one at the spectral minima and an upper state with frequency $\omega_\mathrm{p}-\omega_\mathrm{min}$. Such processes of four-magnon scattering underlie the phenomenon of kinetic instability.
} 
\label{f:KI}
\end{figure}

The increment $\nu_{\bm k}$ becomes positive if
\begin{align}\label{21}
    \begin{split}
        |  \Gamma^{^{\rm KI}}(k) | >   |  \Gamma^{^{\rm KI}}_{\rm th}(k) |= &\frac {\gamma_{\rm top} \gamma_{\rm bot} }{\gamma_{\rm top}+ \gamma_{\rm bot}} \\ \simeq & \gamma_{\rm bot} \,,\quad \mbox{for} \ \gamma_{\rm bot}  \ll \gamma_{\rm top} \ .
    \end{split}
\end{align}
Here, $\gamma_{\rm bot} = \gamma_{2}$ and $\gamma_{\rm top} = \gamma_{3}$ are damping frequencies of the bottom and top magnons, with eigen frequencies around $\omega_\mathrm{min}$ and $\omega_\mathrm{p}-\omega_\mathrm{min}$, respectively.
As the number of parametric magnons ${\cal N}_{\rm par}$ increases and $| \Gamma^{^{\rm KI}}(k) |$ exceeds the threshold value $ | \Gamma^{^{\rm KI}}_{\rm th}(k) | $, kinetic instability typically develops first at the lower end of the frequency spectrum (see Fig.\,\ref{f:KI}), where $\gamma_{\bm k}$ attains its minimum $\gamma_{\rm bot}$.
According to the nonlinear theory of kinetic instability formulated in Ref.\,\cite{Lvov2024}, the exponential growth of the lowest energy magnons is interrupted by their feedback on the parametric magnons.

\subsection{\label{ss:concl}Where we are and the road ahead}
In the previous sections, we theoretically analyzed the parametric pumping regimes and mechanisms for transferring parametrically injected magnons to the bottom of the spectrum, where they can contribute to the formation of a Bose--Einstein condensate.
A change in the pumping conditions affects the spectral distribution of magnons and should alter both the efficiency of each of the considered transport mechanisms and the relationship between them.
However, establishing a direct link between the theories presented in Sec.~\ref{s:exit}, which describe parametric instability under oblique pumping and scattering processes, is not straightforward and is unlikely to be achievable with reasonable effort. 

The primary challenge is that, in order to analytically calculate $\Gamma^{^{\rm KI}}(k)$, which determines the threshold for kinetic instability according to Eq.\,\eqref{20}, one must know the distribution of parametric magnons $n^{\rm par}$ on the resonance surface defined by $\omega_{\bm k} = \omega_{\rm p}/2$. Unfortunately, this distribution does not coincide with the spectral position of the minimal threshold of parametric instability presented in Sec.~\ref{ss:methods}, since the pumping power required for the development of kinetic instability is considerably higher than this threshold \cite{Lavrinenko1981, Kreil2018, Lvov2024}. For significant supercriticality---when the pumping power substantially exceeds the parametric instability threshold---one must consider the redistribution of parametric magnons across the resonance manifold, as described in the Zakharov--Lvov--Starobinets $S$-theory \cite{Zakharov1975, Lvov1993}.

This redistribution results from self-consistent scattering due to phase pairing of parametric magnons, analogous to Bardeen--Cooper--Schrieffer pairing in superconductivity, as detailed in Secs.~5 and 6 of Ref.\,\cite{Lvov1993}. Moreover, Ref.~\cite{Lvov1993} makes it clear that the direct application of the advanced $S$-theory (Sec.~6) requires knowledge of the relevant interaction amplitude $S_{\bm k,\bm k'}$, which has not yet been found for our conditions.

A further complication arises from the fact that calculating $\Gamma^{^{\rm KI}}(k)$ using Eq.\,\eqref{20} requires knowledge of the interaction amplitude $T_{\bm k \bm 1}^{\bm 2\bm 3}$, which is currently unknown. Furthermore, to apply the nonlinear theory of kinetic instability developed in Ref.~\cite{Lvov2024} and to determine the population of bottom magnons under pumping well above the kinetic instability threshold, this amplitude is required again. 
\looseness=-1

Under these circumstances, the only way to determine the optimal conditions for maximizing the number of bottom magnons is to measure ${\cal N}_{\text{bot}}$ over a wide range of applied magnetic fields $\bm H$ and angles between $\bm M$ and the pumping field $\bm h$. The results should then be compared qualitatively with existing theoretical predictions.

\section{\label{sec:exp}Experiment}
This section presents the experimental investigation of parametrically pumped magnons in a YIG film. The setup, described in Sec.~\ref{ss:setup}, employs microfocused Brillouin light scattering (BLS) spectroscopy for magnon detection and includes a microwave pumping circuit for magnon injection, together with a vector magnet that enables continuous in-plane rotation of the external magnetic field. The following subsections present the main results obtained for different pumping geometries. This includes the dependence of the parametric instability threshold on the pumping angle (Sec.~\ref{ss:thresholds}), and angle-resolved BLS measurements revealing the redistribution of magnons toward the spectral minimum (Sec.~\ref{ss:bottom}). The experimental findings are analyzed and discussed in relation to the theoretical framework introduced in Sec.~\ref{s:exit}. \looseness=-1

\subsection{\label{ss:setup}Experimental setup}
In this study, we investigated the parametric pumping and thermalization of magnons in a YIG film with a thickness of \qty{6.7}{\micro\meter} and lateral dimensions of \qty{3.5}{\milli\meter} by \qty{7.5}{\milli\meter}. 
We chose YIG as the experimental magnetic medium because it exhibits the longest known magnon lifetimes, reaching up to \qty{18}{\micro\second} at \qty{100}{\milli\kelvin}~\cite{Serha2025_Ultra-long-living_magnons}.
Also, epitaxial YIG films with micrometer thickness, one of which was used in our experiment, possess magnon lifetimes of \qtyrange{300}{700}{\nano\second} at room temperature~\cite{Serga2010}, significantly exceeding those observed in any other known material.
Such long magnon lifetimes are crucial for their thermalization and the formation of a quasi-equilibrium state, which is necessary for the emergence of a Bose--Einstein condensate \cite{Demokritov2006, Melkov2025}. 

\begin{figure}[b]
\centerline{\includegraphics[width=1\columnwidth]{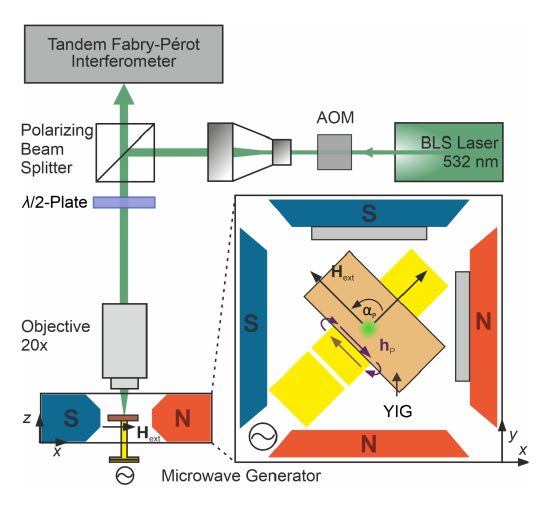}}
\caption{A schematic picture of the experimental setup utilized for the measurements. The magnon spectrum is measured with the micro-focused Brillouin light scattering technique, marked by the green beam path. The inset presents the sample configuration with regard to the external magnetic field $H_{\rm ext}$ and the pumping field $h_{\rm p}$.}
\label{fig:BLS}
\end{figure}

As shown in Fig.~\ref{fig:BLS}, the sample was mounted on a half-wavelength microstrip resonator that had a width of \qty{75}{\micro\meter} and a resonant frequency of \qty{14.094}{\giga\hertz}, which matched the pumping frequency. A microwave signal was delivered to the resonator through a feeding microstrip and a capacitive gap. 
To minimize sample heating caused by high microwave power, we used pulsed pumping with a pulse duration of \qty{4.5}{\micro\second} and a repetition period of \qty{330}{\micro\second}.
 
The assembly was positioned at the center of a vector magnet equipped with four coils, which enabled continuous \qty{360}{\degree} rotation of the external magnetic field $H_\mathrm{ext}$ within the sample plane.
The magnet could produce field strengths of up to $ H_\mathrm{ext}^\mathrm{max} = \qty{2000}{\Oe}$, with a homogeneous region of approximately \qty{1}{\milli\meter\squared} over the sample \cite{Schweizer2023}. 

Magnon detection was performed using microfocused Brillouin light scattering (BLS) spectroscopy~\cite{Sebastian2015, Schweizer2023, Nikolaev2025}, as shown in Fig.\,\ref{fig:BLS}. 
Laser light with a wavelength of \qty{532}{\nano\meter} and power of \qty{20}{\milli\watt} was focused onto the sample using a microscope objective with $20\times$ magnification and a numerical aperture of 0.45, resulting in a laser spot of approximately \qty{3}{\micro\meter} in diameter. 
Incident photons inelastically scattered from magnons in YIG, thereby acquiring a frequency shift equal to the magnon frequency and a \qty{90}{\degree} rotation of polarization. 
The scattered light, whose intensity was proportional to the magnon density, was then directed into a tandem Fabry--P\'erot interferometer~\cite{Mock1987}, which enabled the acquisition of frequency-resolved spectra. 
Turning off the laser beam with an acousto-optic modulator (AOM) during the intervals between the pumping pulses significantly reduced sample heating and prevented a decrease in the density of the condensed magnon state within the focal spot caused by the outflow of magnon supercurrents~\cite{Bozhko2016, Mihalceanu2019, Schweizer2024}. 

Previous studies have indicated that distinguishing between the parallel and perpendicular pumping regimes is not straightforward when detecting the magnon Brillouin light scattering (BLS) signal on a microstrip resonator~\cite{Dzyapko2009}. 
As the pumping field $h_\mathrm{p}$ circulates around the microstrip, as shown in the inset of Fig.~\ref{fig:BLS}, two distinct regions emerge: predominantly in-plane pumping occurs above the resonator, while out-of-plane pumping occurs near its edges \cite{Neumann2009}. However, this spatial inhomogeneity of the microwave field becomes irrelevant when the pumping angle $\alpha_\mathrm{p} = \qty{90}{\degree}$, which corresponds solely to perpendicular pumping. In contrast, for $\alpha_\mathrm{p} = \qty{0}{\degree}$, this effect must be considered, as purely parallel pumping cannot be achieved fully. A key advantage of the micro-focused BLS setup is its small probing spot size relative to the resonator width (approximately $1\!:\!25$), allowing measurements to be taken in the central region of the resonator, where the direction of the microwave magnetic field is well-defined. 

For the threshold measurements characterizing the parametric instability (see Sec.\,\ref{ss:mot}), the pulsed microwave source was replaced by a vector network analyzer. 
This configuration enabled the generation of a quasi-continuous signal delivered to the sample while simultaneously detecting the reflected signal, allowing precise determination of the magnon excitation conditions in the sample \cite{Azevedo2025_Magnonics}. 
 
\subsection{\label{ss:thresholds}Threshold vs external magnetic field for different pumping angles}

The first step toward magnon Bose--Einstein condensation is the realization of overpopulation of the magnon gas, which can be readily achieved by applying an external microwave magnetic field $\bm{h}_\mathrm{p}(t) = \bm{h}_\mathrm{p} \exp(-i\omega_\mathrm{p} t)$. 
As described in Sec.~\ref{ss:mot}, the threshold value of the microwave field $\bm{h}_\mathrm{th}$ depends on the pumping angle $\alpha_\mathrm{p} \equiv \angle(\bm{h}_\mathrm{p}, \bm{H}_\mathrm{ext})$ between $\bm{h}_\mathrm{p}$ and the static magnetic field $\bm{H}_\mathrm{ext}$ (see the inset in Fig.~\ref{fig:BLS}), as well as on the magnitude of $H_\mathrm{ext}$ itself. 
To analyze these dependencies, we measured $h_\mathrm{th}$ for external magnetic fields $H_\mathrm{ext}$ ranging from \qty{1500}{\Oe} to \qty{1900}{\Oe}. 
The pumping angle $\alpha_\mathrm{p}$ was varied from the parallel-pumping geometry ($\alpha_\mathrm{p} = \qty{0}{\degree}$) to the transverse geometry ($\alpha_\mathrm{p} = \qty{90}{\degree}$) in increments of \qty{15}{\degree}. 
As known from previous studies, for a given pumping frequency, the wavevectors of magnons excited under parallel pumping ($\alpha_\mathrm{p}=\qty{0}{\degree}$) undergo substantial changes in both magnitude and direction within this range of magnetic fields~\cite{Neumann2009_2}.
Furthermore, a transition between the regimes of forbidden and allowed kinetic instability, observed in this pumping configuration, significantly affects the efficiency of magnon gas thermalization toward a Bose--Einstein condensate \cite{Kreil2018}.
 
\begin{figure}[t]
\centerline{\includegraphics[width=1\columnwidth]{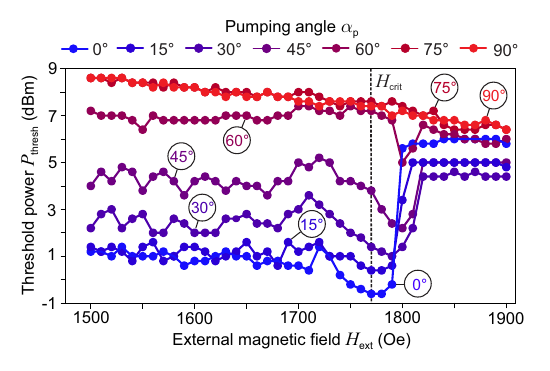}}
\caption{
Threshold power as a function of the external magnetic field for various pumping geometries. 
The characteristic jump observed for parallel pumping ($\alpha_\mathrm{p}=\qty{0}{\degree}$) at the critical field, originating from the leakage of magnons out of the parametric interaction region~\cite{Neumann2009_2}, gradually disappears as $\alpha_\mathrm{p}$ approaches the perpendicular pumping geometry ($\alpha_\mathrm{p}=\qty{90}{\degree}$).
}
\label{f:Threshold}
\end{figure}

The measured threshold curves are shown in Fig.\,\ref{f:Threshold}. 
One can see that the lowest threshold is achieved at $\alpha_\mathrm{p}=\qty{0}{\degree}$ (parallel pumping) near the critical magnetic field $H_\mathrm{crit}$, indicated by the vertical dotted black line. 
At this field, the frequency of the pumped magnons, $\omega_\mathrm{p}/2$, coincides with the minimum frequency of the transverse dispersion branch, $\omega_0 = \omega_{\bm{k}}$, where $\bm{k} \perp \bm{H}_\mathrm{ext}$ and $k = 0$. This magnon state possesses the maximal precessional ellipticity and, therefore, the strongest time-dependent modulation of the longitudinal magnetization, which couples to the pumping field $\bm{h}_\mathrm{p} = h^\parallel$ (see Sec.~\ref{sss:par}). 

For $H_\mathrm{ext} > H_\mathrm{crit}$, the frequency $\omega_0$ exceeds $\omega_\mathrm{p}/2$, which makes the parametric excitation of magnons on the transverse dispersion branch impossible [see Fig.\,\ref{fig:Disp}(b)]. 
Magnons with $k > 0$ excited by parallel pumping on the dispersion branches with $\theta_{\bm k} < \pi/2$ have smaller precessional ellipticities and, consequently, higher thresholds.

Moreover, to realize $\alpha_\mathrm{p}=\qty{0}{\degree}$, the longitudinal axis of the microstrip resonator is oriented perpendicular to the external magnetic field $\bm{H}_\mathrm{ext}$. 
In this geometry, parametric magnons with $\theta_{\bm k}=\pi/2$ excited at $H_\mathrm{ext} < H_\mathrm{crit}$ (see blue dots in Fig.\,\ref{f:KI}) propagate along the several-millimeter-long resonator and interact efficiently with the pumping field $\bm{h}_\mathrm{p}$. 
In contrast, magnons with $\theta_{\bm{k}} < \pi/2$, excited at $H_\mathrm{ext} > H_\mathrm{crit}$, propagate obliquely with respect to the resonator’s longitudinal axis.
Because the resonator itself is only \qty{75}{\micro\meter} wide, these magnons quickly leave the region of parametric interaction due to their high group velocity, leading to a sharp increase in their excitation threshold \cite{Neumann2009_2}. This geometric factor provides an additional, and even stronger, contribution to the increase of the instability threshold compared to the effect of decreasing precessional ellipticity.  

As follows from Sec.~\ref{ss:mot} [see, e.g., Eq.\,\eqref{14}] and as illustrated in Fig.\,\ref{f:KI}, the wavevectors of magnons excited in the transverse pumping process are directed at an angle of $\theta_{\bm k}=\pi/4$ with respect to the external magnetic field. 
Therefore, within the entire range of investigated magnetic fields, these magnons escape from the region of parametric interaction. 
This explains why, for $H_\mathrm{ext} < H_\mathrm{crit}$, their effective damping $\gamma^\perp$, and hence the instability threshold $h_\mathrm{th}^\perp$, are considerably higher than the corresponding values $\gamma^\parallel$ and the threshold $h_\mathrm{th}^\parallel$ of magnons excited by parallel pumping, as indeed observed in Fig.\,\ref{f:Threshold}.

The transition from parallel to perpendicular pumping is accompanied by a monotonic increase in the parametric instability threshold for all magnetic fields $H_\mathrm{ext} < H_\mathrm{crit}$. 
This increase may also be associated with the spatially confined nature of parametric excitation in our experiment.
As the pumping angle $\alpha_\mathrm{p}$ increases, the component of the pumping field parallel to the external magnetic field, $h^\parallel$, decreases.
This reduction lowers the efficiency of parallel pumping even in the absence of magnon outflow or changes in their intrinsic damping.

An interesting question concerns the nonmonotonic variation of the threshold with 
$\alpha_\mathrm{p}$ observed for oblique pumping at $H_\mathrm{ext} > H_\mathrm{crit}$. We assume that this effect arises because the group velocity of parametrically excited magnons, which is perpendicular to the isofrequency contour at $\omega_\mathrm{p}/2$ and generally noncollinear with their wavevector (see, e.g., Refs.~\cite{Pirro2021}), may become aligned with the longitudinal axis of the resonator as shown in Fig.\,\ref{f:KI}. Such alignment reduces the radiation losses of these magnons and, consequently, lowers their instability threshold. Radiation losses may be further modified by the formation of spin-wave caustics in which magnons with different wavevectors propagate in the same direction \cite{Demidov2009, Schneider2010, Papp2018, Heussner2020}.

A detailed comparison of the efficiency of different pumping geometries in exciting magnons, independent of spatial effects, can be achieved by expanding the pumping area using dielectric or cavity resonators. Unfortunately, the pump field strength in such resonators is significantly lower than in microstrip circuits due to the difference in their volumes.
That is why the latter are most often used in magnonics, both in fundamental research \cite{Demokritov2006, Lvov2023, Makiuchi2024, Demokritov2003} and in practical applications \cite{Nikolaev2025, Braecher2017, Kurebayashi2011, Melkov1999}.
Since the objective of our work is to examine the efficiency of thermalization of the pumped magnons toward the bottom of the spectrum, we have limited our study to the most effective and widely used technique of microwave parametric pumping employing a microstrip resonator.

In summary, for all investigated magnetic field magnitudes, the parallel pumping configuration at $\alpha_\mathrm{p} = \qty{0}{\degree}$ exhibits visibly lower threshold values than the corresponding perpendicular pumping configuration, with the difference ranging from 
\qty{7.4}{\dBm} at lower external fields to \qty{0.6}{\dBm} at higher fields. 
Intermediate angles between these two limiting cases correspond to mixed pumping processes. 

\subsection{\label{ss:bottom}BLS measurements of the spectral population}
To compare the pumping efficiencies indicated by the obtained threshold behavior with the resulting population of the magnon spectrum, we performed $\alpha_\mathrm{p}$-dependent BLS measurements. 
The experiment was conducted within the same range of external magnetic fields, from \qty{1500}{\Oe} to \qty{1900}{\Oe}, while maintaining a constant pumping power of \qty{25}{\dBm}. 
To enable a consistent comparison of the measured magnon densities, we introduce the supercriticality parameter $\zeta = \zeta(\alpha_\mathrm{p}, H_\mathrm{ext})$, defined as the ratio of the applied power to the corresponding threshold value \cite{Noack2019}.

\begin{figure}[b]
\centering
\includegraphics[width=1.04\columnwidth]{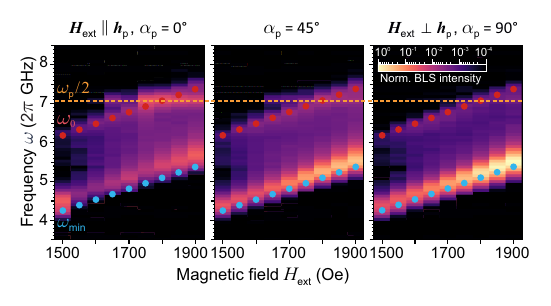}
\caption{ 
Frequency-resolved BLS intensity as a function of the external magnetic field $H_\mathrm{ext}$ 
for three pumping geometries: parallel ($\alpha_\mathrm{p} = \qty{0}{\degree}$), 
oblique ($\alpha_\mathrm{p} = \qty{45}{\degree}$), 
and perpendicular ($\alpha_\mathrm{p} = \qty{90}{\degree}$). 
The magnetic field was incremented in steps of \qty{50}{\Oe}. 
Red and blue dots denote the calculated frequencies of the uniform precession mode ($\omega_0$) and the spectral minimum ($\omega_\mathrm{min}$), respectively, 
while the dashed orange line marks the frequency of parametrically injected magnons 
($\omega_\mathrm{p}/2$).
The BLS intensity was normalized across all three panels to illustrate the intensity difference between the geometries.}
 \label{f:field_sweep}
\end{figure}

Figure\,\ref{f:field_sweep} presents the measurements of the frequency-resolved BLS intensity as a function of the external magnetic field $H_\mathrm{ext}$ for three selected pumping angles $\alpha_\mathrm{p}$: \qty{0}{\degree}, \qty{45}{\degree}, and \qty{90}{\degree}. 
In all three cases, we observe a population of the frequency range between the minimum frequency of the transverse magnon branch, $\omega_0$, corresponding to the frequency of the uniform precession, and the bottom of the spectrum, $\omega_\mathrm{min}$. 
As expected, these limits shift almost linearly upward with increasing external magnetic field. 
For parallel ($\alpha_\mathrm{p} = \qty{0}{\degree}$) and oblique ($\alpha_\mathrm{p} = \qty{45}{\degree}$) pumping, an enhancement of the BLS signal is observed when the frequency $\omega_0$ approaches the frequency of the parametrically excited magnons, $\omega_\mathrm{p}/2$. 
This behavior is explained by the fact that, in these cases, magnons with relatively small wavenumbers are excited, lying within the sensitivity range of our BLS setup. 
In contrast, for perpendicular pumping ($\alpha_\mathrm{p} = \qty{90}{\degree}$), the wavenumbers of parametrically injected magnons are too large to be detected. \looseness=-1

A closer inspection of these data reveals a predominant magnon population near the spectral minimum, most pronounced for the oblique and perpendicular pumping geometries. 
In these two cases, the population of the lowest-energy states increases almost monotonically with the applied magnetic field. 
For parallel pumping, however, this monotonic trend is interrupted by a distinct peak in the magnon density at $H_\mathrm{ext} = \qty{1500}{\Oe}$. 
At this field, the wavevectors of parametrically excited magnons become sufficiently large for the conservation laws to permit direct four-magnon scattering into the spectral minimum \cite{Kreil2018}, indicating the contribution of the kinetic instability process.

\begin{figure*}[t]
\centerline{\includegraphics[width=1\textwidth]{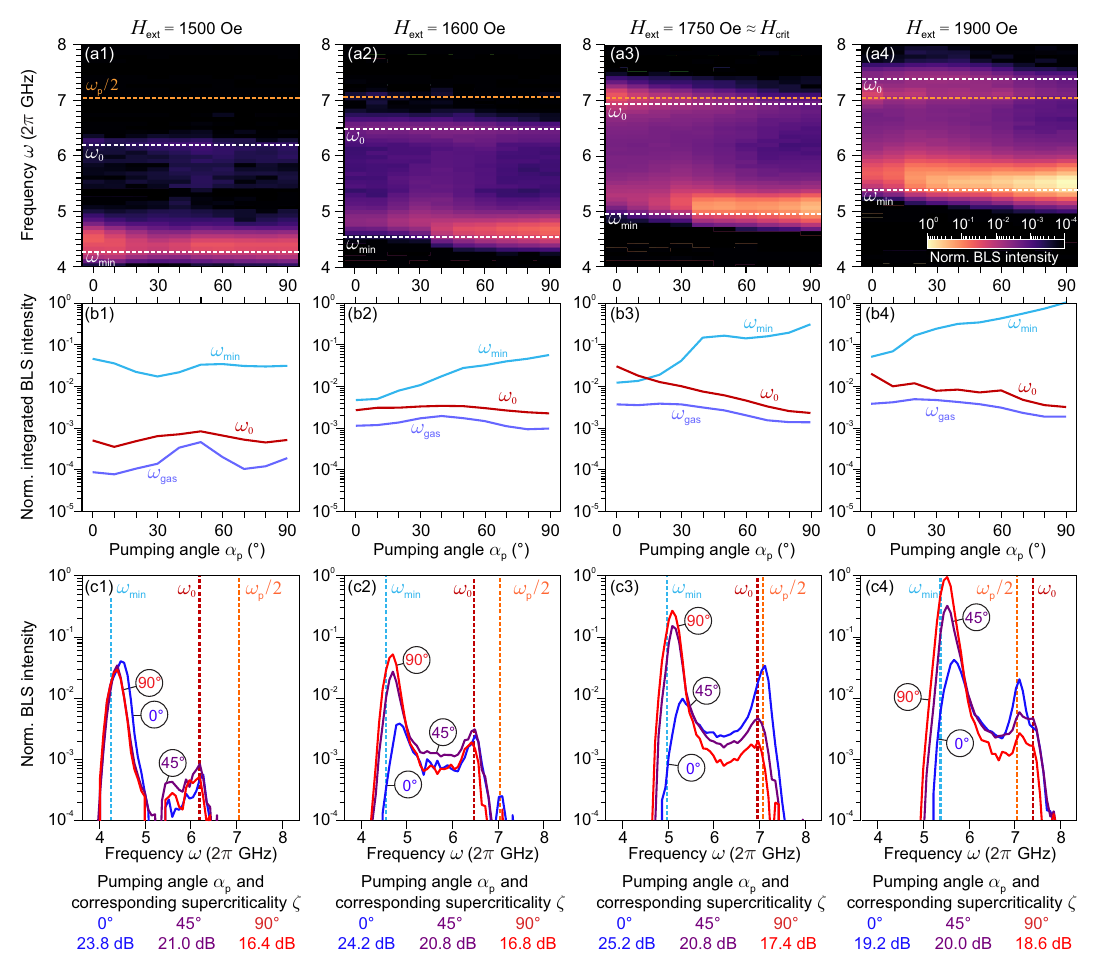}}
\caption{
Frequency-resolved normalized BLS intensity for different external magnetic fields $H_{\rm ext}$:
(a1)~\qty{1500}{\Oe}, (a2)~\qty{1600}{\Oe}, (a3)~\qty{1750}{\Oe}~$\approx H_\mathrm{crit}$, and (a4)~\qty{1900}{\Oe}. 
Each map shows the redistribution of magnons between the frequency of the uniform precession ($\omega_0$) and the spectral minimum ($\omega_\mathrm{min}$) as the pumping angle $\alpha_\mathrm{p}$ varies from parallel ($\qty{0}{\degree}$) to perpendicular ($\qty{90}{\degree}$) geometry.  
The BLS intensities in all panels were normalized to the global maximum value among all measurements.  
(b1)--(b4)~Integrated BLS intensities for three spectral regions: around $\omega_0$, the intermediate gas-like region $\omega_\mathrm{gas}$, and the spectral minimum $\omega_\mathrm{min}$, demonstrating a strong enhancement of the low-frequency population with increasing $\alpha_\mathrm{p}$.  
(c1)--(c4)~Frequency-resolved BLS spectra measured for representative angles $\alpha_\mathrm{p}=\qty{0}{\degree}$, $\qty{45}{\degree}$, and $\qty{90}{\degree}$ under identical pumping power of \qty{25}{\dBm}, with corresponding supercriticality values~$\zeta$ indicated in each panel.  
A pronounced accumulation of magnons at $\omega_\mathrm{min}$ is observed for perpendicular pumping, where the same magnon density is achieved at a 5--6~times lower supercriticality and up to 20--25~times higher population compared to parallel pumping.  
This highlights the crucial role of the pumping geometry in governing the efficiency of magnon transfer toward the Bose--Einstein condensate. 
}
\label{fig:4Fields}
\end{figure*}

Figure\,\ref{fig:4Fields} provides a more detailed view of the magnon spectral population. Figures\,\ref{fig:4Fields}\,(a1)--\ref{fig:4Fields}\,(a4) present the frequency-resolved BLS intensity as a function of the pumping angle $\alpha_\mathrm{p}$ for four selected magnetic fields $H_\mathrm{ext}$: 
\qty{1500}{\Oe} (a1), 
\qty{1600}{\Oe} (a2), 
\qty{1750}{\Oe} (a3), and 
\qty{1900}{\Oe} (a4). 
Each map shows the redistribution of magnons between the frequency of the uniform precession $\omega_\mathrm{0}$ and the spectral minimum $\omega_\mathrm{min}$ as the pumping geometry varies from parallel to perpendicular. 
Here, the BLS intensity in each panel is normalized to the overall maximum value. 

 A pronounced accumulation of magnons at the spectral bottom is clearly visible, and its dependence on the pumping angle varies significantly with the applied magnetic field. 
At the lowest field, where the kinetic instability is allowed, the population of the spectral minimum remains nearly constant over the entire range of pumping angles. 
For the two fields closer to $H_\mathrm{crit}$ [Figs.\,\ref{fig:4Fields}\,(a2) and \ref{fig:4Fields}\,(a3)], the population at small $\alpha_\mathrm{p}$ (i.e., in the regime of forbidden kinetic instability) only slightly exceeds that of the higher-frequency gas-like states. 
At the highest magnetic field, where the kinetic instability at small pumping angles becomes allowed again (see, e.g., Ref.\,\cite{Kreil2018}), the spectral-bottom population at these angles increases to values comparable to those in Fig.\,\ref{fig:4Fields}\,(a1), yet remains substantially lower than that observed for $\alpha_\mathrm{p}$ values approaching the perpendicular-pumping configuration.

This behavior can be evaluated more quantitatively from Figs.\,\ref{fig:4Fields}\,(b1)--\ref{fig:4Fields}\,(b4), which show the integrated BLS intensities derived from the corresponding color maps in panels\,(a1)--(a4). 
For each magnetic field value, the magnon population was integrated in the frequency range $\pm \qty{200}{\mega\hertz}$ over three distinct frequency regions: 
the uniform precession frequency $f_\mathrm{0}$, 
the intermediate gas region $f_\mathrm{gas} = f_\mathrm{min} + \qty{1}{\giga\hertz}$, 
and the spectral minimum $f_\mathrm{min}$.
It should be noted that the integration near the spectral minimum was not performed at a fixed frequency but rather followed the position of the local maximum of the BLS signal.
This approach prevents an artificial underestimation of the magnon density that could occur when the spectral minimum shifts toward higher frequencies.
Such a shift is observed in the parallel-pumping regime, where the threshold of parametric instability is the lowest and, consequently, the number of injected magnons at a fixed power is the highest.
Due to the magnon injection, the saturation magnetization of the YIG film above the resonator decreases. When the external magnetic field is oriented near perpendicular to this elongated area ($\alpha_\mathrm{p} \approx \qty{0}{\degree}$), the internal magnetic field $H_\mathrm{int}$ in the region of reduced magnetization increases due to the stray fields of the surrounding magnetic material. This, in turn, leads to a local rise of the spectral minimum frequency, $\omega_\mathrm{min} = g H_\mathrm{int}$ \cite{Navas2019, Borisenko2020, Schweizer2024}.
 
One can see that the populations of the states with frequencies near the uniform precession frequency $\omega_0$ and those around the intermediate frequency $\omega_\mathrm{gas}$ are comparable for all pumping angles and tend to increase with increasing $H_\mathrm{ext}$. 
Only in cases where parametric magnons are excited by parallel pumping near $\omega_0$, and their wavenumbers are sufficiently small to be efficiently detected by our BLS setup, the population at this frequency becomes noticeably larger. 
The overall increase in the magnon-gas population with growing $H_\mathrm{ext}$ can be attributed to the shift of the well-detectable dipole-exchange spectral region---located between $\omega_0$ and $\omega_\mathrm{min}$---toward the frequency of the parametrically injected magnons, $\omega_\mathrm{p}/2$, resulting in its enhanced occupation. \looseness=-1
 
No significant difference is observed in the population of the spectral minimum at the lowest magnetic field of $\qty{1500}{\Oe}$.
At the same time, as seen from Figs.\,\ref{fig:4Fields}\,(b2)--\ref{fig:4Fields}\,(b4), at higher magnetic fields, the population of the spectral minimum is substantially greater---by up to a remarkable factor of 20--25---in the perpendicular-pumping regime than in the parallel-pumping regime. 
 
To draw conclusions about the efficiency of populating the low-energy states, their occupation must be compared with the corresponding supercriticality values. This comparison is shown in Figs.\,\ref{fig:4Fields}\,(c1)--\ref{fig:4Fields}\,(c4).
They present frequency-resolved BLS spectra measured for the same magnetic fields $H_\mathrm{ext}$ at three representative pumping angles, 
$\alpha_\mathrm{p}=\qty{0}{\degree}$, 
\qty{45}{\degree}, 
and \qty{90}{\degree}. 
These spectra were recorded under equal pumping power of $\qty{25}{\dBm}$, with the different supercriticality values $\zeta$ indicated in each panel for the three pumping angles.  

We find that the same magnon density near the spectral minimum is achieved at a supercriticality that is 5.5 times lower for perpendicular pumping compared to parallel pumping [see Fig.\,\ref{fig:4Fields}\,(c1)]. This indicates a substantially higher efficiency of perpendicular pumping, even under conditions where the kinetic instability is allowed for magnons excited by parallel pumping. 
With increasing magnetic field, when the kinetic instability becomes forbidden, the difference in efficiency becomes striking: the situation shown in Fig.\,\ref{fig:4Fields}\,(c3) corresponds not only to a 6-fold lower supercriticality for perpendicular pumping but also to a 25-fold higher magnon density near $\omega_\mathrm{min}$.
At magnetic fields above $H_\mathrm{crit}$, where the supercriticalities are comparable and the kinetic instability again becomes allowed for magnons excited by parallel pumping, the advantage of perpendicular pumping slightly decreases, although the population of the spectral minimum remains about 20 times higher in this case.
 
\section{\label{s:sum}Discussion and summary}
Our measurements reveal a clear dependence of the magnon spectral population on the pumping angle $\alpha_\mathrm{p}$ between the microwave field $\bm h_\mathrm{p}(t)$ and the external static magnetic field $\bm H_\mathrm{ext}$. 
Although parallel pumping ($\alpha_\mathrm{p} = \qty{0}{\degree}$) exhibits the lowest instability threshold, the resulting population at the spectral minimum remains comparatively small throughout the investigated range of magnetic fields around $H_\mathrm{crit}$. 
This demonstrates that the efficiency of parametric excitation does not directly translate into an efficient transfer of magnons toward the lowest-energy states. 
Instead, the redistribution is governed by the activation of specific four-magnon scattering channels discussed in Sec.~\ref{ss:scattering}. 

For perpendicular pumping ($\alpha_\mathrm{p} = \qty{90}{\degree}$), where the threshold is highest, we observe the opposite tendency: a strong accumulation of magnons near the spectral minimum even at modest supercriticality. 
This behavior indicates that the kinetic-instability mechanism (Sec.~\ref{sss:KI}), which transfers parametrically excited magnons to the bottom of their spectrum in a single step, is generally more efficient than the Kolmogorov--Zakharov step-by-step cascade (Sec.~\ref{sss:Cascade}). 
Consequently, the magnon flux toward the Bose--Einstein condensate is substantially enhanced, yielding up to a $20$--$25$-fold increase of the bottom-state population relative to the parallel-pumping case at comparable conditions (Sec.~\ref{sec:exp}). 

At fields below $H_\mathrm{crit}$, where kinetic instability is allowed for both geometries, the contrast between geometries is reduced. 
Near and above $H_\mathrm{crit}$, when the conservation laws forbid the most favorable parallel-pumping channels, the dominance of perpendicular pumping becomes striking. 
Even at higher fields, where the instability channel for parallel pumping reopens, the bottom-state population achieved under perpendicular pumping remains substantially higher. 
This observation highlights that the perpendicular geometry facilitates a more efficient energy flow toward the lowest magnon states across the entire investigated field range. 

Our measurements provide direct experimental evidence that the pumping geometry enables the selective activation of distinct magnon-scattering pathways, thereby controlling the thermalization flow toward the spectral minimum. 
By adjusting $\alpha_\mathrm{p}$ at fixed power, one can regulate the efficiency of kinetic instability and the resulting bottom-state population (Figs.~\ref{f:field_sweep} and \ref{fig:4Fields}). 
This capability enables the formation of dense, steady-state magnon condensates, paving the way for systematic studies of their nonlinear dynamics, including supercurrents, Josephson phenomena, and self-organized textures. 

In summary, our combined theoretical and experimental study establishes the key role of pumping geometry in determining the efficiency of magnon transfer to the lowest-energy states. 
While parallel pumping minimizes the parametric threshold, perpendicular pumping maximizes the population near $\omega_\mathrm{min}$ due to the enhanced kinetic-instability channel. 
This interplay between injection and nonlinear scattering defines practical routes for controlled generation and manipulation of room-temperature magnon Bose--Einstein condensates. 

\section{Acknowledgments}
This research was funded by the Deutsche Forschungsgemeinschaft (DFG, German Research Foundation) in the framework of TRR 173 -- Grant No. 268565370 Spin+X (Projects B04 and B13).

\appendix
\section{\label{A}Classical Hamiltonian formalism in magnetics}
To determine the thresholds for parallel, perpendicular, and inclined parametric pumping of magnons in ferrodielectrics, we first revisit the Holstein-Primakoff transformation in quantum mechanics\,\cite{Landau2013}
\begin{align}
    \begin{split}\label{HP-tr}
        S_+= S_x+i S_y =& \hbar    \sqrt{2S-  a^\dag a  }\ a \,, \\
        S_+= S_x+i S_y = & \hbar \,  a^\dag  \sqrt{2S- a^\dag a  }\,, \\
        S_z = & \hbar (S - a^\dag a  )\,,
    \end{split}
\end{align}
which expresses the spin operators $S_\pm$ and $S_z$ in terms of the creation and annihilation operators of bosons, $a^\dag$ and $a$. It classical analogue
\begin{align}
    \begin{split}\label{HP-tr1}
        M_+(\bm r, t)= M_x+i M_y =&  b  \sqrt{g (2M_0- g b^*  b  } \,, \\
        M_-(\bm r, t)= M_x-i M_y = &  M_+^*(\bm r, t)\,, \\
        M_z(\bm r, t) = &  (M_0 - b^* b  )\,,
    \end{split}
\end{align}
Here, $g$ is the electron gyromagnetic ratio. In \eqref{HP-tr1} the magnetization vector $\bm M(\bm r, t)$ is expressed in terms of the canonical variables, which are the complex spin-wave amplitudes $b(\bm r, t)$ and $b^*(\bm r, t)$. These amplitudes serve as the classical analogues of the quantum operators $a$ and $a^\dagger$. In this context, $M_0$ represents the saturation magnetization in the absence of spin waves, while $^*$ indicates complex conjugation. 

The equation of motion for $\bm M(\bm r, t)$ was introduced in 1935 by Landau and Lifshitz \cite{Landau1970} (LL). They describe the rotation of the magnetization $\bm M(\bm r, t)$ in response to the effective field $H_{\rm eff}$ and account for not only a real magnetic field but also internal magnetic interactions such as exchange, anisotropy, and dipole-dipole interaction:
\begin{align}
    \begin{split}\label{LLE}
        \frac{d \bm M}{d t}=&  - g \big [\bm M \times H_{\rm eff} ]\,, \\ 
        H_{\rm eff}(\bm r, t)=& \frac{\delta W\{ \bm M(\bm r,t)\} }{\delta M(\bm r, t)}\ .
    \end{split} 
\end{align}
Here, $\delta \dots / \delta M(\bm{r}, t)$ represents functional derivatives, and  $W\{ \bm M(\bm r,t)\}$, a functional of $\bm M(\bm r,t)$, is the energy of the system. The LL dissipation term is skipped here. In a more general form, it will be introduced later. Simple manipulations that lead to \eqref{LLE}
by no means guarantees their correctness \cite{Akhiezer1968}. Following the authors of Ref.\,\cite{Akhiezer1968}, we sweep under the carpet many delicate and complicated problems. Nevertheless, based on our experience in studying the nonlinear behavior of spin waves \cite{Lvov1993}, we believe that the LL Eqs.\eqref{LLE} provide a good first approximation for describing magnons (spin waves). 

As shown, for example in book \cite{Lvov1993}, the LL Eq.\,\eqref{LLE} in variables $b(\bm r, t)$, $b^*(\bm r, t)$ takes the form of a classical Hamiltonian equation
\begin{equation}\label{Heq}
    i \frac{\partial b(\bm r, t)}{\partial t}=\frac{\delta \cal H}{\delta b^*(\bm r, t)} \ .
\end{equation}
The Hamiltonian function, referred to simply as the Hamiltonian, represents the energy $W$ of the system in terms of the variables $ b(\bm r, t) $ and $ b^*(\bm r, t) $.

In space-homogeneous media it is convenient to use instead of $ b(\bm r, t) $ and $ b^*(\bm r, t) $ their Fourier harmonics:
\begin{align}
    \begin{split}\label{FT}
        b(\bm k, t)\equiv & b_{\bm k}=\int b(\bm r, t) \exp (-i \bm k \bm r) \frac{d^3 r}{(2\pi )^{3/2}}\,, \\
        b(\bm r, t)\equiv & b_{\bm r}=\int b(\bm k, t) \exp ( i \bm k \bm r) \frac{d^3 k}{(2\pi )^{3/2}}\ .
    \end{split}
\end{align}
The Fourier transform \eqref{FT} is canonical, meaning that the Hamiltonian Eq.\,\eqref{Heq} retains its canonical form. 
\begin{equation}\label{Heq2}
    i \frac{\partial b(\bm k, t)}{\partial t}=\frac{\delta \cal H}{\delta b^*(\bm k, t)} \ .
\end{equation}
Here Hamiltonian $\cal H$ is the functional of $ b(\bm k, t) $ and $ b^*(\bm k, t) $. 

Now is the time to recall that in the LL Eq.\,\eqref{LLE}, we skipped the damping term that is proportional to some constant. Instead, we added the phenomenological damping term $\gamma(\bm k)$ in Eq.\,\eqref{Heq2}, which needs to be clarified later, either experimentally or theoretically: 
\begin{equation}\label{Heq3}
    i\Big [  \frac{\partial  }{\partial t}+ \gamma(\bm k)   \Big ]b(\bm k, t) =\frac{\delta \cal H}{\delta b^*(\bm k, t)} \ .
\end{equation}
The expansion of the Hamiltonian $\mathcal{H}$ for small amplitudes starts with the quadratic terms: $ \mathcal{H}= \mathcal{H}_2 + \dots $, where 
\begin{align}
    \begin{split}\label{H2}
        \mathcal{H}_2 = \int \Big \{ A_{\bm k}b_{\bm k} b_{\bm k}^* + \frac 12 \big [B_{\bm k}b_{\bm k} b_{-\bm k} + \mbox{c.c}\big] \Big\}d^3 k\ .
    \end{split}
\end{align}
Here ``c.c.'' refers to the complex conjugate. The Hamiltonian \eqref{H2} can be diagonalized by a linear Bogoliubov  canonical $u-v$ transformation 
\begin{subequations}\label{uv}
    \begin{eqnarray}\label{uvA}
        b_{\bm k}&=& u_{\bm k} c_{\bm k}+ v_{\bm k} c_{-\bm k}^*\,,\\ \label{uvB}
        \mathcal{H}_2 &=& \int \omega_{\bm k}c_{\bm k} c_{\bm k}^* d^3 k \ .
    \end{eqnarray}
\end{subequations}
In variables $c_{\bm k,t}$ the Hamiltonian Eq.\,\eqref{Heq3} with ${\cal H}={\cal H}_2$, given by Eq.\eqref{uvB} becomes trivial:
\begin{equation}\label{A9}
    i\Big [\frac {\partial}{\partial t}+\gamma_{\bm k}\Big ] c_{\bm k}= \frac{\delta {\cal H}_2}{\delta c_{\bm k}^*} = \omega_{\bm k}c_{\bm k} \ .
\end{equation} 
It has solution the $c(\bm k,t) =  c(\bm k,0)\exp [ - (\gamma _{\bm k} + i\omega _{\bm k})t ] $
which describes the free propagation of a spin wave with frequency $\omega _{\bm k}$ and damping decrement constant $\gamma_{\bm k}$. 

The next terms of the Hamiltonian expansion $ \mathcal{H} = \mathcal{H}_2 + \mathcal{H}_3 +\mathcal{H}_4 $ describe interactions of three and four magnons. 
In the presence of an external homogeneous microwave field, one also has to account for the so-called ``pumping'' Hamiltonian, ${\cal H}^\angle_{\rm p}$ given by Eqs.\,\eqref{9}.

\bibliography{Enhancement_of_magnon_flux}

\end{document}